\documentclass{aa}  

\usepackage{graphicx}
\usepackage{txfonts}
\usepackage{hyperref}
\usepackage[dvipsnames]{xcolor}
\usepackage{booktabs}
\usepackage{siunitx}
\usepackage[normalem]{ulem}
\usepackage{orcidlink}

\newcommand{\gaia}{\textit{Gaia}\xspace}

\defcitealias{2024A&A...692A..96L}{L24}

\begin{document} 

   \title{The case for an Astrometric Mission Extension of \textit{Euclid}} 
   \subtitle{Extending \textit{Gaia} by 6 magnitudes with \textit{Euclid} covering one-third of the sky}

\titlerunning{Extending \textit{Gaia} with \textit{Euclid}}
\authorrunning{Bedin}
%

   \author{
           \texttt{Luigi ``Rolly'' BEDIN}\inst{1}\orcidlink{0000-0003-4080-6466}\fnmsep\thanks{\email{luigi.bedin@inaf.it}}
         }

   \institute{Istituto Nazionale di Astrofisica (INAF), Osservatorio Astronomico di Padova, Vicolo dell’Osservatorio 5, Padova I-35122, Italy\\
             }

   \date{Received 25 September 2025 / Accepted 27 October 2025}


\abstract{
The nominal duration of \textsc{Euclid}'s main mission is six years, but current best estimates indicate that the observatory has sufficient propellant to operate for up to $\sim 14$ years in total. 
In this work, we advocate dedicating six of these $\sim 8$ additional years 
to repeating the main survey, covering approximately one-third of the sky.
 This repetition would not only improve the sampling, signal-to-noise, quality, and depth of the survey, but---most importantly---would provide a six-year time baseline between two epochs if executed in the same sequence. The availability of multiple epochs would enable the derivation of proper motions for stars as faint as $V \approx 27$, i.e., more than five magnitudes fainter than those measured by the \textsc{Gaia} mission. Although it may seem early to propose such a mission extension, in this work we quantitatively illustrate its immense scientific potential. We therefore intend to initiate the technical and scientific discussions early to ensure optimal planning. The here proposed extension would employ only the VIS channel---owing to its superior astrometric capability and depth---while simultaneously using NISP in slitless-spectroscopy mode to enhance the signal-to-noise ratio of first-epoch spectra that would also benefit of proper motions to identify and reject objects within the local Universe.
}


\maketitle

%
\section{Introduction}
%

Over the past $\sim$30\,years, astronomers worldwide have benefited from the unparalleled capabilities of the \textit{Hubble Space Telescope (HST)}: its high angular resolution, stability, and the remarkable quality of its images obtained from the pristine darkness of space. However, these images have been restricted to relatively few and narrow fields of view, and the possibility of observing the entire sky at an \textit{HST}-comparable resolution has remained, until now, only a dream.  

\begin{figure*}[th!]
    \centering
    \includegraphics[width=14truecm]{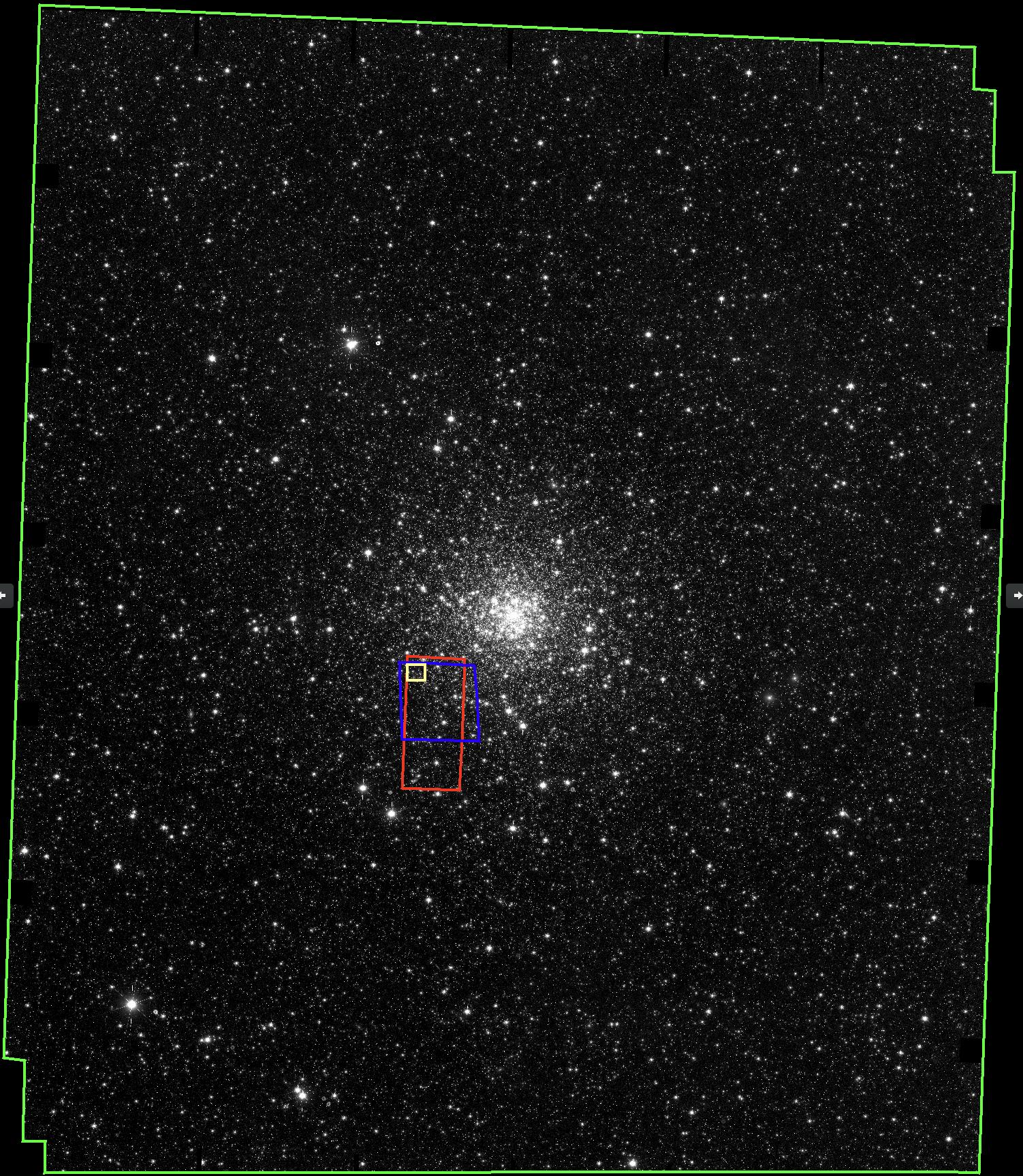}
    \includegraphics[width=14truecm]{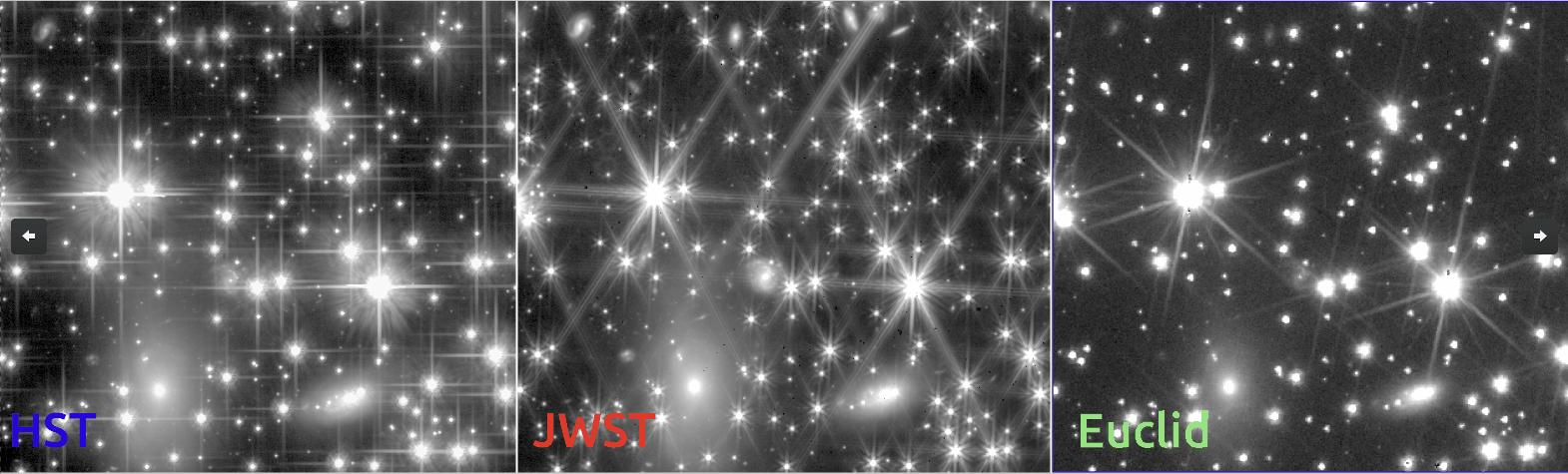}
\caption{Comparison of imaging depth and resolution from different space observatories. 
On top, \textit{Euclid} VIS observations of the globular cluster NGC\,6397 as combined by \cite{2024A&A...692A..96L}. 
While both \textit{HST} and \textit{JWST} achieve substantially greater imaging depths than \textit{Euclid} in these specific data sets (\textit{HST} large program GO-10424 and \textit{JWST} program GO-1979), 
the angular resolution of \textit{Euclid} is comparable to that of \textit{HST} and \textit{JWST}, enabling high-precision morphological studies over a considerably larger survey area. 
The \textit{Euclid} $\sim1^{\circ}\!\times\!1^{\circ}$ field of view presented in \citetalias{2024A&A...692A..96L} is shown in green, 
the \textit{JWST} $\sim2.5'\!\times\!6'$ field studied by 
\citet[program GO-1979]{2021jwst.prop.1979B,2024AN....34540039B} 
in red, 
and the \textit{HST} $\sim3'\!\times\!3'$ field from program GO-10424 in blue. 
The yellow region, of about $50''\!\times\!50''$, in the top panel is common to the three data sets, which are shown individually in the bottom panels with observatories labelled according to the same colour code.
[Images and catalogues available here: 
\href{https://web.oapd.inaf.it/bedin/files/PAPERs_eMATERIALs/JWST/GO-1979/P01/}{\textit{HST \& JWST}} and 
\href{https://web.oapd.inaf.it/bedin/files/PAPERs_eMATERIALs/Euclid/Paper01/}{\textit{Euclid}}]. 
}
    \label{fig:HST-JWST-Euclid}
\end{figure*}

With the advent of the \textit{Euclid} space observatory 
\citep{2025A&A...697A...1E}
we are closer than ever to realising  this vision, as \textit{Euclid}, enables observations of about one-third of the sky at a resolution and depth similar to those of \textit{HST} (Fig.\,\ref{fig:HST-JWST-Euclid}).  

High resolving power directly translates into high-precision imaging astrometry. However, in order to maximize sky coverage while minimizing both the number of pixels (i.e., telemetry requirements) and readout noise, wide-field imagers are often under-sampled, and \textit{Euclid} is no exception. Its detectors range from moderately to severely under-sampled 
\citep{2024CuillardPhot,2024A&A...692A..96L}.

A comprehensive discussion of the challenges in achieving unbiased astrometric precision in under-sampled images from space-based observatories, and the methods to recover the full astrometric information, is provided in the seminal work of \citet[hereafter AK00]{2000PASP..112.1360A}. This study introduced an iterative procedure to break the degeneracy and to solve empirically and simultaneously for both the \textit{effective} point-spread functions ($e$PSFs) and the positions of sources.  

\begin{figure*}[th!]
    \centering
    \includegraphics[width=\textwidth]{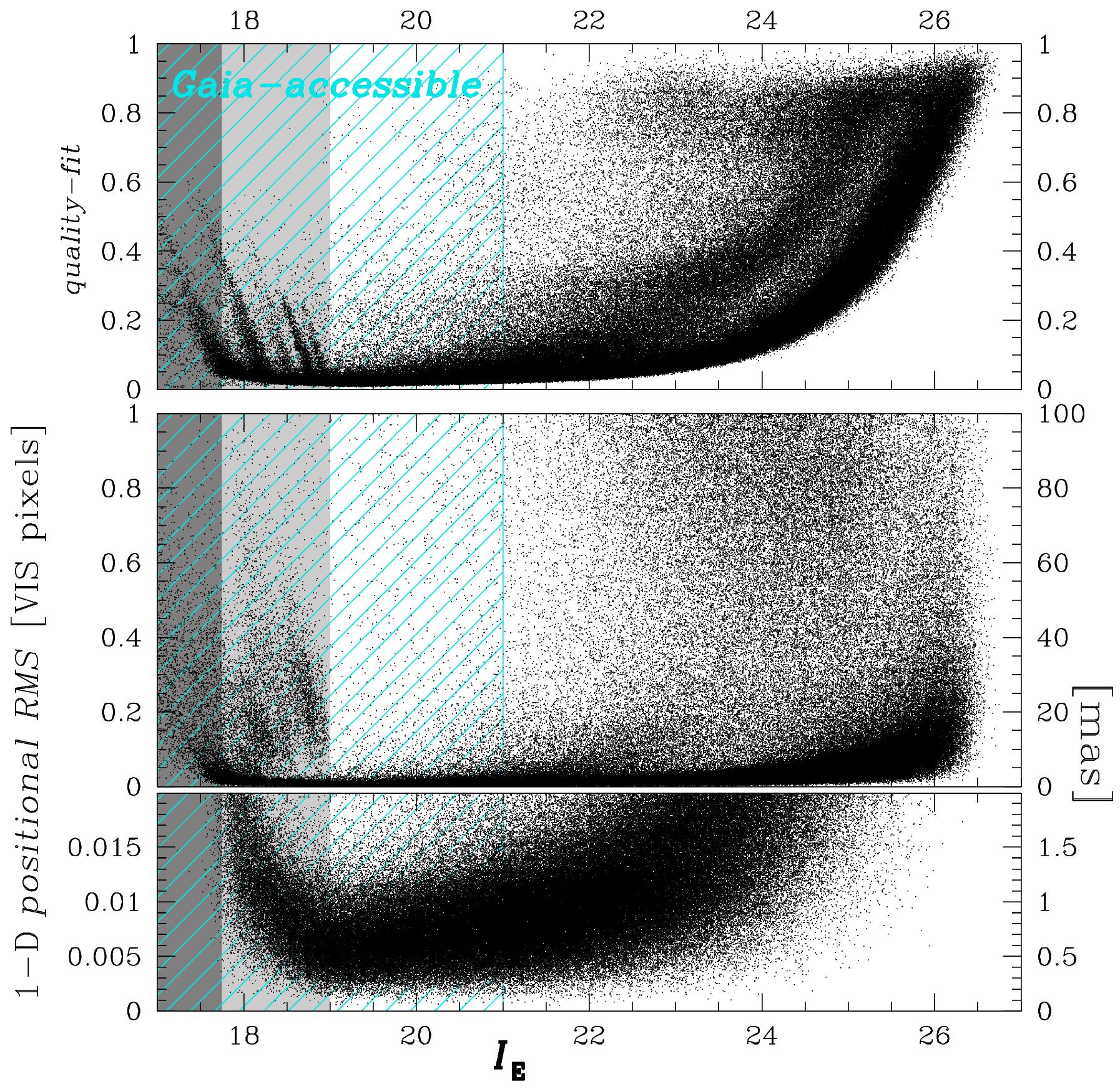}
    \caption{
    The published catalogue by \cite{2024A&A...692A..96L} 
    have  demonstrated the 1D-positional RMS precision of VIS@\textit{Euclid} imaging astrometry.
    Mild saturation begins at $I_{\rm E} \sim 19$  (light-gray shaded regions), while sources brighter than 
    $\sim$18 are severely saturated (region in dark-gray). 
    The region of the sources accessible to \textit{Gaia} is indicated by a shaded region in azure. 
    \textit{(Top:)} Shows the \textit{quality-fit} parameter \citep[see][this parameter quantifies 
    how well a source profile resemble the PSF model]{2024A&A...692A..96L} as 
    function of the calibrated VIS-magnitude in filter $I_{\rm E}$. 
    \textit{(Middle:)} Illustration of the astrometric precision as function of the magnitude 
    (in VIS pixels on left axis, and in mas on the right axis). 
    \textit{(Bottom:)} 
    The precision is as good as 20 mas in a single exposure for sources at $I_{\rm E} \sim 26$, 
    but it can reach sub-mas precision on a single image for a well-exposed unsatured point sources. 
    }
    \label{fig:rmsall}
\end{figure*}

In a recent study, \citet[hereafter L24]{2024A&A...692A..96L} demonstrated that, once undersampling is properly treated, \textit{Euclid}'s detectors are capable of delivering astrometric positional precisions down to the sub-milliarcsecond (sub-mas) level for unsaturated, high signal-to-noise sources, and as good as 10\,mas for sources as faint as $V\sim27$
(see Fig.\,\ref{fig:rmsall}).  
~\citetalias{2024A&A...692A..96L} also described the rather complex procedures required to fully recover the astrometric signal in \textit{Euclid} images. The lack of optimal calibration data to independently solve for both the $e$PSFs and the geometric-distortion corrections (GDCs) further complicates the procedures, which are necessarily empirical and iterative.  

\medskip
Astrometric positions from the single-epoch \textit{Euclid} catalogue already enable, and will continue to enable, important synergies and diverse scientific applications when combined with the \textit{Gaia} catalogues, as partly discussed 
in \citetalias[see][]{2024A&A...692A..96L}. 
Notably, \textit{Euclid} images reach substantially fainter magnitudes 
than \textit{Gaia} --- up to approximately six magnitudes deeper, depending on the spectral energy distribution of the sources.

In this work we propose dedicating part of the (expected) extra operational years beyond the nominal \textit{Euclid} mission to repeating as much of the already surveyed sky as possible, with a second epoch separated by about six years. This strategy would enable the determination of proper motions (PMs) for sources much fainter (and, naturally, far more numerous 
on a same patch of sky) than those available in \textit{Gaia} DR3 or in any future \textit{Gaia} data release.  

The author is member of the \textit{Euclid} Consortium. This article, however, is written in his capacity as independent researcher. The proposals and views expressed herein are solely those of the author and should not be regarded as representing the official position of the \textit{Euclid} Consortium or its governing bodies.

\medskip
This paper is organized as follows. 
We briefly outline in Sect.\,\ref{S:euc} \textit{Euclid} mission and its stellar astrophysics capabilities. 
In Sect.~\ref{S:obs} we describe the typical \textit{Euclid} observing strategy for most sky regions
(the reference observation sequence, ROS), and how real-data imaging astrometry of 
these compare with the available and future \textit{Gaia} catalogues. 
In Sect.~\ref{S:second_epoch} we outline the proposed new observations within the mission extension, and use the preliminary astrometric and photometric performance of \textit{Euclid} (as presented in \citetalias{2024A&A...692A..96L}) to make quantitative predictions for the resulting catalogues. 
In Sect.~\ref{S:two} we explore the concept of collecting two extra epochs instead of just one. 
In Sect.~\ref{S:sci} we present a brief overview of the scientific applications enabled by this added legacy value of the \textit{Euclid} mission. 
In Sect.\,\ref{S:proscons} we summarize the advantages and disadvantages of collecting additional epochs with the \textit{Euclid} mission. 
Finally, Sect.~\ref{S:end} provides our conclusions.  


%
\section{Point sources and stellar astrophysics}\label{S:x}
\label{S:euc}
%

The \textit{Euclid mission}, launched on July 1, 2023, is a joint ESA--NASA project designed to explore the nature of dark energy and dark matter by mapping the geometry of the Universe. Operating from the Sun--Earth L$_2$ point, \textit{Euclid} is conducting a six-year survey of roughly 15,000 square degrees of the sky, capturing high-resolution optical images and near-infrared spectra of over 1.5 billion galaxies and about 30 million spectroscopic redshifts 
\citep{2025A&A...697A...1E}. 
Equipped with two main instruments—the VISible imager (VIS) and the Near-Infrared Spectrometer and Photometer (NISP)—\textit{Euclid} aims to reconstruct the large-scale structure of the cosmos and understand its evolution over the past 10 billion years. While its primary focus lies in cosmology, \textit{Euclid}’s unprecedented data is already having significant impacts on stellar astrophysics, the study of our Galaxy’s structure and evolution, and our knowledge of the Milky Way’s satellite systems.

\textit{Euclid}’s early results, particularly from the \textit{Early Release Observations} (ERO), 
demonstrate its immense power for stellar astrophysics 
\citep[e.g.:][and references therein]{esa2024ero,2025A&A...701A..40G,2025A&A...697A...6C,2025A&A...697A...9H}. 
%
These single-day observations revealed a breathtaking diversity of objects: star-forming regions, open and globular clusters, nearby galaxies, galaxy clusters, and diffuse stellar populations. \textit{Euclid} cataloged over 11 million objects in visible light and about 5 million in infrared, enabling astronomers to probe stellar populations at unprecedented scales. Among its notable findings is the detection of numerous \emph{free-floating planets} and \emph{brown dwarfs}—objects that bridge the gap between stars and planets—suggesting that the Galaxy may host hundreds of thousands of such substellar bodies, possibly down to Saturn-like masses. 
\citep{martin2025, esa2024ero}.
These discoveries are crucial for understanding stellar formation, planetary system evolution, and the role of low-mass objects in Galactic dynamics.

\textit{Euclid}’s high-resolution imaging also provides fresh insights into the \textit{structure and evolution of the Milky Way}. While not primarily designed as a Galactic survey, its wide-field coverage captures stars across the disk, bulge, and halo, complementing missions such as \emph{Gaia} and \emph{JWST}. By characterizing stellar populations and tracing their motions and compositions,  \textit{Euclid} helps build a clearer picture of how the Milky Way assembled its mass over cosmic time and how its structure fits into the broader context of galaxy evolution within the cosmic web. Moreover, by mapping galaxies across cosmic history, \textit{Euclid} indirectly informs models of how galaxies like the Milky Way formed, interacted, and evolved in various environments, including groups and clusters.

One of \textit{Euclid}’s most transformative contributions concerns the \textit{satellites of the Milky Way and nearby dwarf galaxies}. The ERO data revealed a ``rich harvest'' of previously undetected faint dwarf galaxies, improving our census of low-mass companions around massive galaxies \citep{marleau2025ero, esa2024press}. 
In particular, \textit{Euclid}’s deep imaging of galaxy clusters, such as Perseus, enabled measurements of the luminosity and stellar mass functions down to extremely faint magnitudes ($M \approx -11.3$). Intriguingly, these observations suggest fewer low-mass galaxies than predicted by standard cosmological simulations, posing challenges to our understanding of dark matter distribution and dwarf galaxy formation. Such findings have direct implications for solving long-standing problems in astrophysics, like the ``missing satellites'' problem, and help refine models of how small galaxies form, survive, or are disrupted in the environments around larger systems like the Milky Way.

In essence, \textit{Euclid} is far more than a cosmological probe. While its overarching goal is to unveil the nature of dark energy and dark matter, its wide-field, high-resolution view of the Universe is transforming multiple areas of astrophysics simultaneously. By uncovering populations of rogue planets and substellar objects, mapping the structure and evolution of our Galaxy, and expanding our knowledge of faint dwarf galaxies and satellite systems, \textit{Euclid} is reshaping our understanding of how the Milky Way fits into the grand narrative of cosmic history. Its continuing surveys promise an unprecedented synergy between cosmology and stellar astrophysics, offering a new, holistic view of our place in the Universe.

%
\section{\textit{Euclid} Work-Horse Observing Strategy}
\label{S:obs}
%
%
The \textit{Euclid} mission employs a carefully designed observing strategy to achieve its primary scientific objectives of probing the nature of dark energy and dark matter through precise measurements of galaxy shapes and redshifts across a large portion of the extragalactic sky. Central to this strategy are the Wide Survey, which covers approximately 15,000~deg$^2$, and the Deep Fields, totaling about 50~deg$^2$, supplemented by Deep Auxiliary Fields of 6.5~deg$^2$ that provide ultra-deep imaging for calibration and specialized scientific studies  
(see Figures\,\ref{fig:EuclidsSky}). 
Observations are conducted using the VISible Imaging Channel (VIS) and the Near-Infrared Spectrometer and Photometer (NISP) operated (mostly) simultaneously. Each pointing in both cameras employs a four-point dithering pattern with an offset of roughly 100~arcseconds, designed to ensure uniform coverage, improve sampling of the point spread function, fill gaps between detectors, and facilitate rejection of cosmic ray and artifacts (such as hot-, warm-pixels, snow-balls, persistence, etc...).

%
For the Wide Survey, the data collection at each of the four pointings follow 
the reference observation sequence (ROS), which we can summarize in two steps \citep{2022A&A...662A.112E}:
\begin{itemize}
    \item \texttt{step-1}: the nominal exposure time for VIS filter $I_E$ is 566\,seconds per pointing, while NISP performs \textit{simultaneous} 
              spectroscopic observations with grisms lasting 549.6\,seconds per exposure. For a total duration with overheads 
              for this step of 585\,s.\\ 
    \item \texttt{step-2}: Once this step is concluded, NISP-only is operated to collect 
            imaging in the $Y_E$, $J_E$, and $H_E$ bands with 87.2\,seconds per exposure. In this second part of the sequence 
            VIS \textit{is not} collecting (deep) images because of lacking of telemetry and of potential vibrations introduced by the 
            mechanical filter-change of NISP. The total duration of this second phase is 388\,s. 
        \\  
%
\end{itemize}
The dithering strategy is optimized to balance uniform sky coverage with spacecraft operational constraints, so that each sky position is observed multiple times and on different parts of the detectors, allowing both systematic effects and instrumental variations to be mitigated. The Deep Fields employ a more concentrated observing pattern, with multiple exposures per field to achieve greater depth, reaching fainter magnitudes and enabling high-resolution (and accuracy) calibration for the Wide Survey.

Sky coverage proceeds according to a Reference Survey Sequence (RSD) that accounts for calibration requirements, spacecraft pointing constraints, and the distribution of bright sources and background light across the sky (see Figures\,\ref{fig:EuclidsSky}). Observations are scheduled to optimize efficiency, minimize gaps, and maintain uniform depth while balancing visibility and thermal constraints of the spacecraft. Calibration observations are interleaved with science exposures and include both photometric and spectroscopic checks. 
In addition to the main science exposures, \textit{Euclid} periodically acquires short calibration exposures aimed at monitoring the stability of the VIS and NISP instruments and at validating the spectroscopic performance of the survey. These observations are used to assess potential temporal drifts in instrument response, ensure accurate PSF and photometric calibration, and verify the completeness and purity of the redshift catalogue by comparing against reference fields with known spectroscopic redshifts \citep{2022A&A...662A.112E}.

Prior to the main mission, Early Release Observations (ERO) were conducted over 24 hours on a set of 17 astronomical targets including galaxy clusters, nearby galaxies, globular clusters, and star-forming regions. These observations employed the same four-dither strategy used in the main surveys, with VIS exposures in filter $I_E$ of 566\,seconds and NISP exposures of 87.2\,seconds in the $Y_E$, $J_E$, and $H_E$ bands, demonstrating the capabilities of \textit{Euclid} and providing early public data. By combining wide and deep observations, precise dithering, and careful calibration, the \textit{Euclid} mission ensures high-fidelity imaging and spectroscopy, allowing it to map the large-scale structure of the Universe with unprecedented accuracy and to advance our understanding of dark energy and dark matter.

%
\subsection{Comparison of \textit{Euclid} and \textit{Gaia}\,DR3 Astrometry}
\label{S:VSgaia}
%
%
%
To illustrate the power of \textit{Euclid}’s imaging astrometry, 
in this section we compare its results with those of the best \textit{all-sky} astrometric catalogue currently available 
--- the \textit{Gaia} source catalogue in its most recent public release, DR3.
This comparison, discussed in part in \citetalias{2024A&A...692A..96L}, is presented here to quantitatively demonstrate 
\textit{Euclid} real data, not to diminish in any way the extraordinary scientific value of the \textit{Gaia} catalogues.
For clarity, we divide this comparison into three magnitude regimes (see Fig.\,\ref{fig:rmsall} for reference):
%
\footnote{
For simplicity, we adopt a crude approximation by assuming that stars saturating in \textit{Euclid} images 
at a magnitude of $I_{\rm E}\!\sim\!19$ correspond to a \textit{Gaia} magnitude of $G\!\sim\!19$. 
In reality, this relation depends strongly on the spectral energy distribution (SED) of each source. 
Likewise, we assume that the faintest stars detected by \textit{Gaia}, are at $G\!\sim\!21$, correspond to $I_{\rm E}\!\sim\!21$.
}

\medskip
\noindent
\textbf{\textit{Gaia} Domain --- $G\lesssim19$:}  
Most sources with $G<19$ in the \textit{Gaia} catalogues are saturated, or nearly saturated, in \textit{Euclid} images. 
Their point-spread functions (PSFs) ---and therefore their astrometric positions--- are significantly affected. 
For these bright sources, \textit{Euclid} cannot improve or complement the astrometric precision or accuracy 
of any past or future \textit{Gaia} data releases.
%

\medskip
\noindent
\textbf{\textit{Euclid} Domain --- $G\gtrsim21$:} 
At the faint end, most sources with $G>21$ are simply too faint to be detected or measured by \textit{Gaia} 
and are therefore absent from DR3 and will remain so in all future \textit{Gaia} releases.
In this regime, \textit{Euclid}’s astrometric catalogue becomes invaluable, providing positional 
measurements for sources up to six magnitudes fainter than those included in \textit{Gaia}.
%

%

\noindent
\textbf{Synergy Domain --- $19\lesssim G \lesssim 21$:} 
This intermediate range deserves special attention. 
In this magnitude interval, sources are measurable by both \textit{Gaia} and \textit{Euclid}. 
The high signal-to-noise \textit{Euclid} positions can contribute to better constraining the astrometric solutions derived 
from the fit to the individual \textit{Gaia} measurements of these faint stars.\\
More importantly, even a single-epoch \textit{Euclid} catalogue enables the derivation of \textit{Gaia}-\textit{Euclid} proper motions (PMs), as demonstrated in \citetalias{2024A&A...692A..96L}. 
By combining \textit{Euclid} positions (epoch  $\sim$2023.7) with \textit{Gaia} DR3 positions (epoch 2016.0), 
high-precision proper motions (PMs) can be obtained. 
These combined PMs were shown to improve the accuracy of \textit{Gaia}\,DR3 values by up to an order of magnitude.\\~ 
%

%
\noindent
However, while \textit{Gaia}-\textit{Euclid} PMs based on DR3 show improvements by up to a factor of ten, 
this advantage will naturally diminish as newer \textit{Gaia} releases become available. 
The upcoming \textit{Gaia}\,DR4 (expected December 2026) will likely reduce this factor to about 4–5, 
and DR5 to roughly 2 for the majority of the stars in the synergy domain \citep{2024eas..conf..208B}. 
Nevertheless, we note that improving the proper-motion precision by a factor of two (in mas\,yr$^{-1}$) increases the accessible volume of space by a factor of eight, for a given target precision in 
tangential velocity (in km\,s$^{-1}$).

\begin{figure*}[th!]
    \centering
    \includegraphics[width=\textwidth]{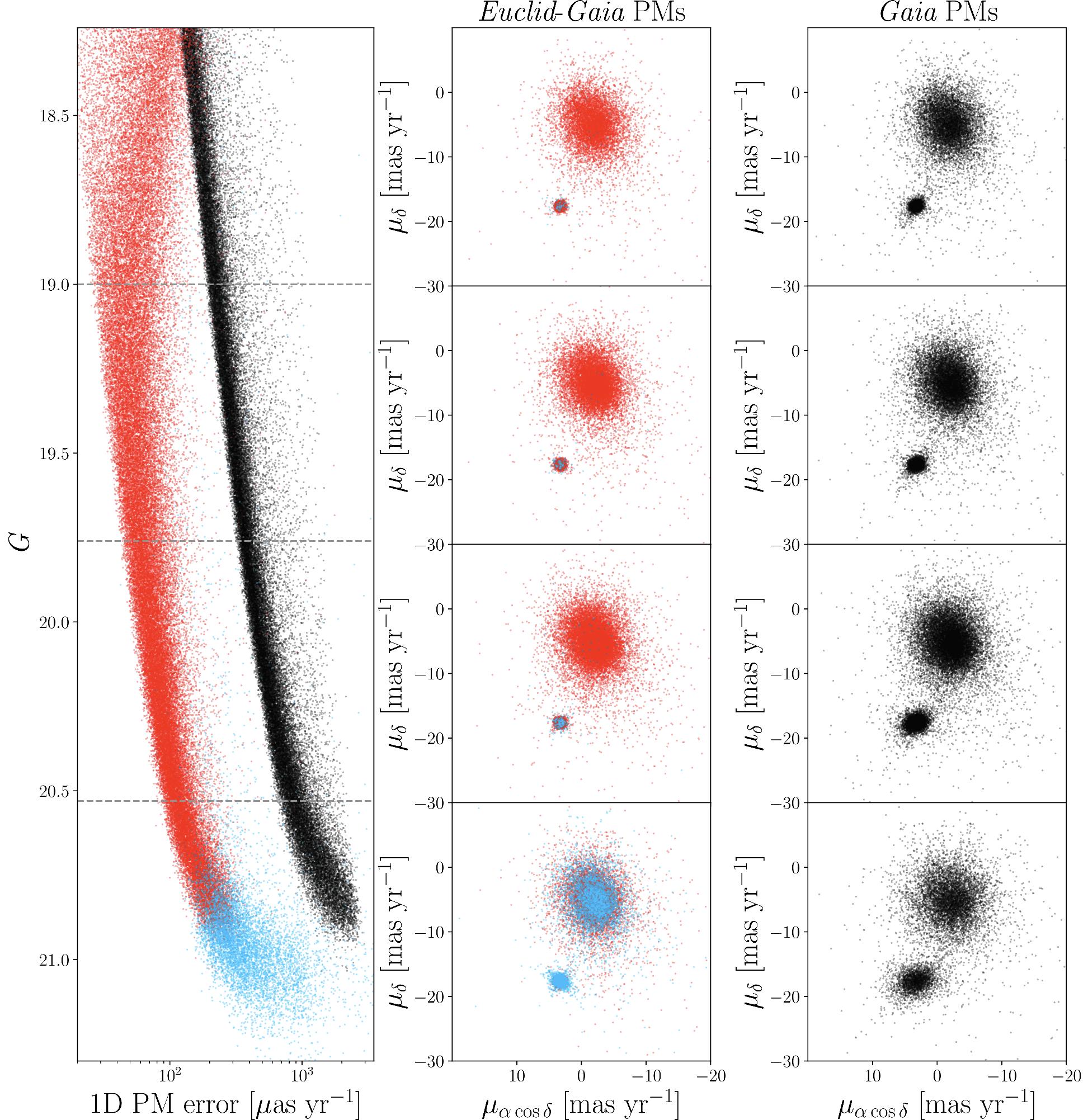}
    \caption{
    Overview of the comparative astrometric analyes of \textit{Euclid} and \textit{Gaia} as presented by \citetalias{2024A&A...692A..96L}. The left panel shows \gaia-$G$ magnitude as a function of 1D PM error (the sum in quadrature of the PM errors along the $\alpha\cos\delta$ and $\delta$ directions divided by $\sqrt 2$). Stars present in both the \gaia DR3 and \textit{Euclid}-\textit{Gaia} catalogs are shown in black and red, respectively, while blue points are sources with a PM measurement only in the \textit{Euclid}-\textit{Gaia} catalog (i.e., with positions but no motions available in the \textit{Gaia}\,DR3 catalogues). The vector-point diagrams of the PMs in four magnitude bins (the thresholds of which are drawn in the left panel) are shown in the middle and right panels for the \textit{Euclid}-\textit{Gaia} and \gaia DR3 cases, respectively. Clearly \textit{Euclid-Gaia}\,DR3 proper motions are ten-times more precise in this common magnitude interval, and provide PMs for sources with positions entries in \textit{Gaia}\,DR3, but not PMs.
    However, the significant improvements over DR3 seen in the narrow magnitude range $19 < G < 21$ are expected to become progressively smaller with the advent of the upcoming DR4 and DR5 of far superior 
    astrometric precision and accuracy (see text).
    }
    \label{fig:EuclidVsGaiaDR3}
\end{figure*}

\medskip
In addition, proper motions can be derived for sources that currently lack them in DR3 
---those with only two-parameter (\texttt{2p}) astrometric solutions, i.e., positions only. 
The comparison between \textit{Gaia}\,DR3-\textit{Euclid} PMs and \textit{Gaia}\,DR3 PM values, 
as well as the newly obtained PMs for \texttt{2p}-only sources, are shown in Fig.\,\ref{fig:EuclidVsGaiaDR3}, 
based on the \citetalias{2024A&A...692A..96L} catalogue.

\medskip
It must also be clearly stated and emphasized that the link between unsaturated \textit{Euclid} sources and \textit{Gaia} entries 
provides the astrometric anchoring of \textit{Euclid} data to an absolute reference system. 
Because \textit{Euclid} covers only relatively narrow fields, it cannot independently determine absolute positions 
except with respect to background galaxies ---whose centroids are less reliable due to morphological complexity and 
deviations from point-source PSFs.\\~
Finally, future \textit{Gaia} releases will include the individual epoch astrometric measurements for each source. 
The combination of these data with \textit{Euclid}’s high signal-to-noise positions for faint stars in this magnitude range 
will further improve the fit of all the astrometric parameters (positions, PMs and parallaxes) of common objects. 
%

%
\section{A Second Epoch of the \textit{Euclid} Wide Survey}
\label{S:second_epoch}
%
%
In this work, we advocate for conducting a \textit{second epoch} of the entire \textit{Euclid} Wide Survey (\textit{E}WS) 
--- or as large a fraction of it as operationally feasible --- to begin immediately after the nominal six-year duration of the main mission. The second epochs should aim at repeating 
the \texttt{step}-1 (described in Sect.\,\ref{S:obs}) of the survey in a manner as identical as possible to the original, including the same cadence, 
observing strategy, dither, and instrument configuration. 

Aspirationally, we envision repeating the entire $\sim$15,000 deg$^2$ of the \textit{E}WS, though even a partial re-observation of selected regions would yield significant scientific benefits. From an operational standpoint, we foresee no major technical obstacles to this plan. 
Current estimates \citep{2024eas..conf.2675G}\footnote{
\texttt{https://eas.unige.ch/EAS2024/plenary.jsp}. Public and private communication.
}
suggest that the spacecraft’s propellant should allow for up to an additional eight years of operations after the nominal six-year mission, which would enable a full re-observation of the same sky area. 
Furthermore, planning a second epoch limited to just the ``\texttt{step}-1'' of the \textit{Euclid} observing sequence (described in previous section), 
would shorten the observation to $\sim$60\% of the \texttt{step}-1+\texttt{step-2} duration of the main mission, making the repeating of the survey faster, or 
potentially offering the collection of extra images and spectra per pointing. 

A second epoch of the \textit{E}WS, carried out at equal depth, would provide a remarkable scientific return, far exceeding the alternative of extending the survey to cover an additional one-third of the sky at shallower depth. By design, the already-surveyed area represents the darkest regions available; hence, re-observing the same fields would maximize sensitivity while boosting photometric signal-to-noise. The scientific advantages of this approach are manifold: not only would it reinforce the primary \textit{Euclid} science goals, but it would also enable a wealth of new opportunities in both extragalactic and stellar astrophysics.\\ 

One of the most compelling motivations for a second \textit{Euclid} epoch 
is the ability to measure \textit{proper motions}  for an unprecedented number of 
sources; fainter than what reachable by \textit{Gaia} (i.e., $G$\,$\gtrsim$21). 
Taking into account the updated source statistics from the most recent estimates presented at the 2024\,EAS meeting \citep{2024eas..conf..208B}, the total number of objects in the \textit{Gaia} catalogue is expected to 
increase from about 1.8\,billion in DR3 to nearly 2.8\,billion in DR4, with roughly one billion of 
them being \texttt{2p} sources  --- that is, entries with only positional information. 
In the subsequent DR5 release, these numbers will grow only moderately. 
Since \textit{Euclid} will survey approximately one-third of the sky, 
it will overlap with of order $3\times10^8$ of the \textit{Gaia} \texttt{2p} sources 
and roughly $5\times10^8$ of the \texttt{5p+} sources.\\ 
When considering the same area of sky, \textit{Euclid} is expected to detect roughly five times more sources 
than those contained in \textit{Gaia}\,DR3 (\citetalias{2024A&A...692A..96L}),  
owing to its $\sim6$\,mag deeper limiting sensitivity. 
However, this ratio will naturally decrease as future \textit{Gaia} releases include more faint sources. 
Based on current forecasts \citep{2024eas..conf..208B}, the \textit{Euclid}-to-\textit{Gaia} 
source ratio will drop from 
$\sim$5$\times$ relative to DR3, to about 
3$\times$ for DR4, and to roughly 
2.5$\times$ for DR5, 
still representing a substantial increase in the number of detectable objects over the same sky area. 
Even with these updated figures, \textit{Euclid} will therefore remain a unique and complementary facility, 
extending the reach of space-based astrometry to several magnitudes fainter than the final \textit{Gaia} catalogues.
\\
%

Based on current estimates, a time baseline of $\sim$6--8 years between the two epochs would allow proper-motion uncertainties down to the few 100~$\mu$as\,yr$^{-1}$ level at the \textit{E}WS detection limit. 
These unprecedented astrometric capabilities would enable clean separation between Galactic and extragalactic sources, improve Milky Way structural models, trace stellar streams and tidal debris in the halo, and provide new constraints on the formation and evolution of the Local Group (more in Sect.\,\ref{S:sci}).\\

In addition to astrometry, a second epoch would have profound consequences for the \textit{slitless spectroscopy} obtained by the Near Infrared Spectrometer and Photometer (NISP). Doubling the exposure time on every target would directly improve the signal-to-noise ratio of \textit{Euclid}’s low-resolution spectra, enabling more robust redshift measurements, especially for faint galaxies near the current detection limit. This deeper spectroscopic dataset would reduce catastrophic redshift failures, improve clustering analyses, and enhance constraints on cosmological parameters. Moreover, fainter emission-line galaxies would become accessible, broadening the redshift range and increasing the number density of spectroscopic tracers available for cosmological studies.

Also as benefit to cosmology and for the \textit{Euclid}’s primary mission, additional epochs would enhance the characterization of the GDC and of the $e$PSF, refining our understanding of their fine structures and of their temporal and spatial variations. This, in turn, would improve the accuracy of weak-lensing measurements. Moreover, proper-motion data could be exploited to distinguish unresolved background \textit{fix} galaxies from foreground stellar objects \textit{moving} within the Local Group.

In summary, a second epoch of the \textit{E}WS would transform \textit{Euclid} from a static imaging and spectroscopic mission into a \textit{multi-epoch astrometric powerhouse}, extending its reach well beyond its initial science goals. The combination of doubled imaging depth, enhanced spectroscopy, and precise proper motions for billions of sources would open a vast landscape of new scientific opportunities, as further discussed in Sect.\,\ref{S:sci}.

%
\section{Contemplating Two Additional Epochs}
\label{S:two}
%
%
In the nominal mission plan, the 
\textit{Euclid} Wide Survey 
provides the first epoch (hereafter \emph{epoch\,1}) over a baseline of six years. In Sect.~\ref{S:second_epoch} we advocated the significant scientific benefits of obtaining a \emph{second epoch} of the \textit{E}WS, repeating the survey as identically as possible to the main mission. Here, we consider an
even more ambitious scenario: collecting \textit{two additional epochs} (\emph{epoch\,2} and \emph{epoch\,3}) after the completion of the nominal mission.

Acquiring two extra epochs would naturally be more challenging from both an operational and resource standpoint, but the potential scientific rewards are considerable. In particular, with three opportunely well-separated epochs, \textit{Euclid} would be able to measure \textit{stellar parallaxes} for billions of stars, opening up a new and unique window into Galactic structure and stellar astrophysics. Crucially, to achieve meaningful parallax sensitivity, the three epochs must not all be observed at the same time of the year: at least one of the additional epochs must be obtained approximately six months out of phase with respect to one of the others, in order to sample the parallactic ellipse optimally.

In the simplest scenario, the two extra epochs would match the exposure times and observing strategy of the first epoch, thus doubling the observational investment relative to the single-epoch case discussed in Sect.~\ref{S:second_epoch}. However, one could consider a compromise approach: collecting only \emph{half the exposure time} in each of the new epochs, effectively acquiring two pointings per field instead of four. While this would lead to shallower imaging, it would still provide the necessary astrometric leverage for parallax measurements when combined with the deep first epoch (or a further compromised choice, e.g., three pointings few months apart).  

The feasibility of acquiring observations at phases separated by six months depends on a number of \textit{engineering and operational constraints} that require detailed assessment by the \textit{Euclid} operations team \citep{2022A&A...662A.112E}. 
Specifically, it needs to be evaluated whether the spacecraft can safely and efficiently 
reorient \citep[][Sect.\,2.3]{2022A&A...662A.112E} its observing strategy to sample the desired parallactic angles, while maintaining thermal stability, power constraints, and mission efficiency. If such an approach proves technically feasible, it would allow \textit{Euclid} to deliver parallax measurements for sources up to $\sim$6 magnitudes fainter than the \emph{Gaia} limit. Thanks to the common-unsaturated magnitude interval with \emph{Gaia} 
all the \textit{Euclid} astrometric measurements will be in an \textit{absolute reference system}, dramatically improving astrometric precision and accuracy \citep[e.g.][]{2018MNRAS.481.5339B,2020MNRAS.494.2068B}. 

Therefore, while repeating the main mission strategy for one identical second epoch is straightforward and operationally viable, extending to two additional epochs spaced by six months introduces non-trivial engineering challenges. Nevertheless, the potential science gains --- combining deep \textit{Euclid} photometry with high-precision parallaxes for billions of faint sources --- are so significant that this possibility warrants serious consideration and further discussion within the \textit{Euclid} collaboration and the astronomical community.

\medskip
As discussed at the end of Sect.\,\ref{S:VSgaia}, the additional astrometric data points provided by \textit{Euclid} 
will contribute to the fitting of \textit{Gaia}'s individual astrometric measurements, 
thereby tightening the constraints on all astrometric parameters at the faint end of the \textit{Gaia} catalogue. 
Indeed, starting with \textit{Gaia}\,DR4, the individual 
observations will be publicly released, 
allowing for joint solutions that combine \textit{Gaia} and \textit{Euclid} measurements. 
This will enable the derivation of accurate parallaxes and proper motions for a large fraction of the currently two-parameter (\texttt{2p}) sources, in much the same way as the early \textit{Gaia}\,DR1 data were combined 
with \textit{Hipparcos} and \textit{Tycho} measurements to improve astrometric solutions. 
In this context, even a single additional \textit{Euclid} epoch will substantially enhance the precision of astrometric parameters for faint stars, while a third \textit{Euclid} epoch ---offset by about six months--- would provide valuable redundancy for parallax determinations and secure reliable solutions for the faintest sources that might otherwise remain 
poorly constrained in \textit{Gaia}~DR5 plus \textit{Euclid-epoch\,2} alone.
%

%
\section{Scientific Opportunities}
\label{S:sci}
%

The addition of one or more extra epochs to the 
\textit{Euclid} Wide Survey 
would open a vast landscape of new scientific opportunities, far beyond the mission’s original cosmological objectives. By doubling the number of VIS images (and NISP spectra) in each field and enabling time-domain astrophysics, multi-epoch \textit{Euclid} observations would deliver unprecedented capabilities in stellar astrophysics, Galactic archaeology, extragalactic science, and transient phenomena, while simultaneously improving the main \textit{Euclid} science goals such as weak gravitational lensing and galaxy clustering. Below, we summarize some of the most promising science cases.

\subsection*{Stellar Astrophysics and Galactic Archaeology}
\begin{itemize}
    \item \textbf{Demography of low-mass stars and brown dwarfs}: Deep multi-epoch imaging will probe the faint end of the stellar luminosity function in the disk, halo, and bulge, as well as the population of brown dwarfs (BDs) in the Galactic field. Proper motions derived from repeated observations will enable robust separation between nearby substellar objects and distant extragalactic sources, providing a complete census of cool, low-mass populations down to planetary-mass objects.
    \item \textbf{Faint and ultracool white dwarfs}: A temporal baseline will improve detection of very faint, cool, and massive white dwarfs, critical tracers of the oldest stellar populations. The resulting statistics will inform models of stellar evolution, star-formation history, and Galactic chemical enrichment.
   \item \textbf{Stellar streams, halo substructure, and accretion history}: Combining precise proper motions with deep photometry allows the detection of faint tidal streams, disrupted satellites, and other low-surface-brightness structures in the Galactic halo, thereby constraining the merger history of the Milky Way and the properties of dark matter. According to current survey forecasts, the 
\textit{E}WS
   will cover approximately $\sim$25 known stellar streams and $\sim$15 significant halo substructures within its footprint, while also providing sensitivity to detect many more ultra-faint streams predicted by $\Lambda$CDM. The addition of a second or third epoch would enable accurate mapping of their kinematics, revealing the signatures of past accretion events and constraining the Galactic gravitational potential on large scales. Multi-epoch \textit{Euclid} data will thus provide an unprecedented census of tidal debris and faint satellites, placing powerful constraints on dark matter models and the hierarchical assembly history of the Milky Way.    
    \item \textbf{Star clusters and internal dynamics}: Two or three epochs would allow kinematic measurements of stars within globular clusters and open clusters, probing their internal velocity dispersion, mass segregation, and evaporation rates. Within its current footprint, the \textit{Euclid} Wide Survey is expected to observe approximately $\sim$25--30 known globular clusters and about $\sim$250 open clusters, spanning a broad range of distances, metallicities, and dynamical states. In combination with parallaxes and accurate photometry, this will enable robust determinations of dynamical masses, tidal radii, and structural parameters for these systems. Moreover, the ability to measure proper motions for stars several magnitudes fainter than the \emph{Gaia} limit will allow us to probe cluster outskirts in unprecedented detail, revealing tidal tails, ongoing disruption, and the effects of the Galactic tidal field on cluster survival. This opens unique insights into the interplay between cluster evolution, the Galactic potential, and the early assembly history of the Milky Way.
    \item \textbf{Dwarf galaxies and ultra-faint satellites}: Thanks to its wide-area coverage, exquisite angular resolution, and sensitivity to low surface brightness, the \textit{Euclid} Wide Survey will provide an unprecedented census of nearby dwarf galaxies. Within its $\sim$15,000$\,\mathrm{deg}^2$ footprint, \textit{Euclid} is expected to re-observe nearly all of the $\sim$60 known Milky Way satellites and to discover at least $\sim$80$-$120 new dwarf galaxy candidates, extending the census to $M_V \sim -6$ and down to surface brightness levels as faint as $\mu_V \sim 31 \,\mathrm{mag\,arcsec^{-2}}$. Furthermore, \textit{Euclid} will cover more than $70\%$ of the known nearby galaxy groups, enabling the detection of hundreds of ultra-faint dwarf galaxies out to distances of $\sim$5\,Mpc. This will significantly improve constraints on galaxy formation in low-mass halos and provide a direct test of $\Lambda$CDM predictions for the abundance of dark-matter-dominated systems.
\end{itemize}

\noindent
While two epochs would provide unprecedented proper motions, the addition of a \textbf{third epoch}, optimally offset in phase by about six months, would also enable the measurement of \textbf{stellar parallaxes} down to magnitudes as faint as $V \sim 26$. Access to accurate distances would provide direct determinations of \emph{absolute magnitudes}, placing individual stars and substellar objects on well-calibrated luminosity functions. This capability would allow us to break degeneracies between intrinsic luminosity and distance, enabling more robust determinations of the mass function of low-mass stars and brown dwarfs, the ages and metallicities of faint white dwarfs, and the three-dimensional structure of stellar streams, clusters, and tidal debris. In this way, parallaxes from a third epoch would provide crucial additional constraints on stellar astrophysics and Galactic archaeology, significantly enhancing the scientific return of the \textit{Euclid} mission.

\subsection*{Time-Domain Science and Transients}
Multiple epochs of \textit{Euclid} will be 
particularly effective in revealing variabile sources 
considering the simultaneous surveys by 
observatories such as \textit{LSST} and the upcoming \textit{Roman Space Telescope}. 
\begin{itemize}
    \item \textbf{Supernovae and other transients}: Multi-epoch imaging naturally enables the discovery and characterization of supernovae, kilonovae, tidal disruption events, and other variable phenomena. \textit{Euclid}’s relatively-red filter $I_E$ 
    would be particularly powerful for high-redshift supernovae and dust-obscured events.
    \item \textbf{Stellar variability and exotic objects}: With photometric monitoring over multiple years, \textit{Euclid} can detect variability across a wide range of stellar types, including flare stars, pulsating variables, eclipsing binaries, and rare compact objects such as cataclysmic variables or AM CVn systems.
    \item \textbf{Microlensing events}: The repeated coverage of the same fields at high angular resolution could allow serendipitous detection of microlensing events by compact objects, including possible constraints on isolated stellar remnants and even free-floating black holes. Once a luminosity variation is detected and flagged, these candidate events could be promptly followed up with other space- and ground-based facilities, enabling spectroscopic characterization, multi-band photometric monitoring, and precise modelling of the lensing geometry.
\end{itemize}

\subsection*{Enhancing \textit{Euclid}’s Core Science}
\begin{itemize}
    \item \textbf{Improved PSF modeling and weak lensing performance}: Doubling the number of images per field would enable temporal modelling of both the \textit{Euclid} Point Spread Functions and of the Geometric Distortion Correction (\citetalias{2024A&A...692A..96L}), improving control of systematics in weak gravitational lensing analyses.
    \item \textbf{Better photometric redshifts and spectroscopy}: Doubling the exposure time of slitless NISP spectra would directly improve the signal-to-noise ratio, leading to more accurate and complete redshift estimates, especially for faint galaxies near the current detection limit.
    \item \textbf{Higher-precision galaxy catalogs}: Multi-epoch imaging would enable construction of deeper, cleaner (decontaminated from foreground moving stars), and more complete galaxy and point-source catalogs, benefiting a wide range of astrophysical studies.
\end{itemize}

\subsection*{Extended Astrometric Capabilities}
\begin{itemize}
    \item \textbf{Proper motions and parallaxes}: With two epochs separated by $\sim$6--8 years, \textit{Euclid} can derive proper motions for up to 
    3\,billions sources \citepalias{2024A&A...692A..96L}, reaching $\sim$6 magnitudes fainter than the \textit{Gaia} limit. With three epochs, strategically offset in orbital phase by six months, parallaxes would become feasible even at very faint magnitudes, enabling unprecedented three-dimensional mapping of the Milky Way and its satellites.
    \item \textbf{Synergy with \textit{Gaia} and \textit{LSST}}: 
A powerful way to obtain precise light curves in crowded fields is to use an external high-resolution catalogue to model and subtract neighbouring stars via the instrument’s PSF. This neighbour-subtraction technique reduces blending and yields cleaner light curves than aperture photometry which is always affected by dilutions effects \citep[e.g., ][and references therein]{2019MNRAS.490.3806N,2015MNRAS.447.3536N,2016MNRAS.456.1137L}. An \textit{Euclid}-based input list, with its depth and angular resolution, would be an ideal reference for LSST, enabling accurate photometry in dense regions and extending its discovery potential for faint variables, transients, and exoplanet signals.
\end{itemize}

\subsection*{Legacy Science Value}
The combined imaging, spectroscopy, astrometry, and time-domain information from multi-epoch \textit{Euclid} observations will provide a transformative, multi-purpose dataset. The resulting catalogs of point sources, parallaxes, proper motions, galaxy morphologies, and spectral properties will represent a unique resource for the community, supporting investigations across virtually all fields of astrophysics for decades to come.

\section{Pros and Cons of Additional Epochs}
\label{S:proscons}

Adding one or more extra epochs to the \textit{Euclid} Wide Survey comes with a range of scientific benefits but also technical and operational challenges. Below we summarize the main considerations.

\subsubsection*{Scientific Advantages}
\begin{itemize}
    \item \textbf{Dramatic improvement in astrometry}: A second epoch would enable proper motion measurements for up to $\sim$3\,billion stars down to VIS $\sim$26\,mag, roughly 6 magnitudes fainter than \emph{Gaia}, with precision of few $\sim$100~$\mu$as level for the faintest sources. A third epoch, suitably offset in phase, would additionally enable parallax determinations for faint stars, providing an unprecedented 3D map of the Milky Way and its satellites.
    \item \textbf{Boost in photometric depth and S/N}: Doubling the exposure time for both VIS images and NISP spectra leads to improved signal-to-noise ratios and more effective rejection of artifacts, better photometric redshifts, and higher-quality slitless spectra, particularly for faint galaxies near the current detection limit.
    \item \textbf{Enhanced PSF modeling and weak lensing}: More epochs improve the temporal modeling of the PSF, benefiting weak lensing shape measurements and reducing systematics. A better characterization of GDC will also enhance calibration accuracy.
    \item \textbf{Time-domain astrophysics}: Additional epochs enable the discovery and characterization of variable sources, transients, tidal disruption events, supernovae, microlensing, and other rare phenomena.
    \item \textbf{Legacy value}: Multi-epoch \textit{Euclid} data would create an unparalleled astrophysical dataset, complementing \emph{Gaia}, \emph{LSST}, and \emph{Roman}, enabling a vast range of community-driven science for many years. 
\end{itemize}

\subsubsection*{Technical and Operational Challenges}
\begin{itemize}
    \item \textbf{Charge Transfer Efficiency (CTE) degradation}: Over time, radiation damage in space causes CTE losses in the VIS CCD detectors, degrading astrometric and photometric performance. However, lessons from \emph{HST/ACS-UVIS} and \emph{Gaia} show that CTE effects can be accurately modeled and corrected even after 10--20 years of operation, meaning this is a manageable, not prohibitive, issue.
    \item \textbf{Thermal stability and instrument aging}: Long-term stability of the optical system and thermal control is critical for maintaining PSF quality and astrometric accuracy. While small drifts are expected, they can be mitigated through recalibration strategies.
    \item \textbf{Mission planning and fuel budget}: While current estimates \citep{2024eas..conf.2675G} predict sufficient propellant for up to $\sim$8 years of extended operations, operational constraints, scheduling priorities, and data volume limitations must be carefully evaluated.
    \item \textbf{Telemetry and data processing}: Doubling or tripling the number of exposures increases the data volume significantly, requiring enhanced capabilities and additional resources for calibration, reduction, and archiving.
    \item \textbf{Engineering feasibility of non-synchronous epochs}: If parallaxes are desired, one of the additional epochs must be observed $\sim$6 months out of phase relative to the others. This raises complex scheduling and thermal constraints that require dedicated assessment by the \textit{Euclid} operations team.
\end{itemize}

\subsubsection*{Lessons from \textit{HST} and \textit{Gaia}}
Experience from long-duration space missions like \emph{HST} (still operating successfully for more than 30 years) and \emph{Gaia} (2013-2025) demonstrates that detector aging, CTE degradation, and optical stability challenges are surmountable with proper calibration pipelines. \textit{Euclid} benefits from this accumulated expertise and is well-positioned to maintain high-quality astrometry and photometry even beyond its nominal mission lifetime.

Overall, while some technical hurdles exist, none appear fundamentally prohibitive. The scientific return from one or more additional epochs would be transformative, significantly amplifying \textit{Euclid}'s legacy and enhancing its synergies with future surveys.

%
\section{Conclusions}
\label{S:end}
%
%
The \textit{Euclid} mission already represents a transformative leap forward in high-resolution imaging and cosmological mapping, but its potential to deliver groundbreaking astrometric science is far from fully realized. Leveraging its exceptional stability, wide-field imaging, and sub-milliarcsecond astrometric precision, \textit{Euclid} can extend the reach of \textit{Gaia}’s legacy by more than six magnitudes, accessing billions of fainter sources across one-third of the sky. 

By dedicating a portion of the mission’s anticipated extended lifetime to conducting a second epoch of the \textit{Euclid} Wide Survey,  
we could unlock unprecedented scientific opportunities. Two well-separated epochs would enable accurate proper motions for up to 
$\sim$3\,billion stars, reaching down to $V \sim 26$\,mag \citepalias[estimates based on results by][]{2024A&A...692A..96L}. 

If technically feasible, adding a \textit{third epoch}, optimally offset by six months in orbital phase, would represent 
a further transformative leap, enabling precise \textit{parallax measurements} down to $V \sim 26$. This would extend three-dimensional mapping of the Milky Way and its satellite systems into regimes unreachable by any existing or planned mission, revolutionizing our understanding of Galactic structure, stellar populations, and the assembly history of the Local Group.  

Beyond astrometry, the science benefits are multifaceted. A second epoch would double the depth of \textit{Euclid}’s slitless spectroscopy, significantly improving redshift completeness and accuracy, expanding access to faint emission-line galaxies, and enhancing clustering analyses and cosmological constraints. 

Furthermore, repeated imaging has the potential to produce significant advances in time-domain astrophysics, encompassing phenomena such as supernovae, microlensing events, variable stars, and tidal streams, with implications spanning stellar, Galactic, and extragalactic regimes. The high-angular-resolution input catalog and detections provided by \textit{Euclid} will facilitate accurate and precise characterizations in ongoing surveys conducted by observatories such as LSST and Roman, with methods similar to those described in 
\cite[][and reference there in]{2019MNRAS.490.3806N}. 

Therefore, even a modest extension of the \textit{Euclid} mission would transform it into a multi-epoch astrometric powerhouse, extending \textit{Gaia}’s impact by an order of magnitude in both depth and scope. Such an investment would pay extraordinary dividends across a vast range of science cases, ensuring \textit{Euclid}’s enduring legacy in mapping the Universe and our place within it.\\ 

Finally, the long-term value of a multi-epoch \textit{Euclid} survey would be immense. The resulting catalogs -- comprising precise positions, parallaxes, proper motions, high-quality spectra, and morphological measurements -- would form a legacy dataset of unparalleled depth and breadth, complementing \emph{Gaia}, \emph{LSST}, and the \emph{Roman Space Telescope}. 
This synergy would secure European leadership in space-based astrometry well into the coming decades and provide the foundation  for addressing some of the most pressing questions in astrophysics and cosmology.

\begin{acknowledgement}
The author is grateful to 
the referee Dr.\, Ulrich (Uli) Bastian 
for the prompt and thorough evaluation of the manuscript. 
The constructive comments ---particularly those concerning the use of current and forthcoming \textit{Gaia} catalogues--- have been invaluable in improving both the clarity and the overall quality of the paper. 
The author also thanks Dr., Mattia Libralato for carefully reading the manuscript and for his insightful suggestions.
\end{acknowledgement}


\bibliography{EE}{}
\bibliographystyle{aa}

\begin{appendix}

\section{Euclid survey field of view}
\begin{figure*}[th!]
    \centering
    \includegraphics[width=18.2truecm]{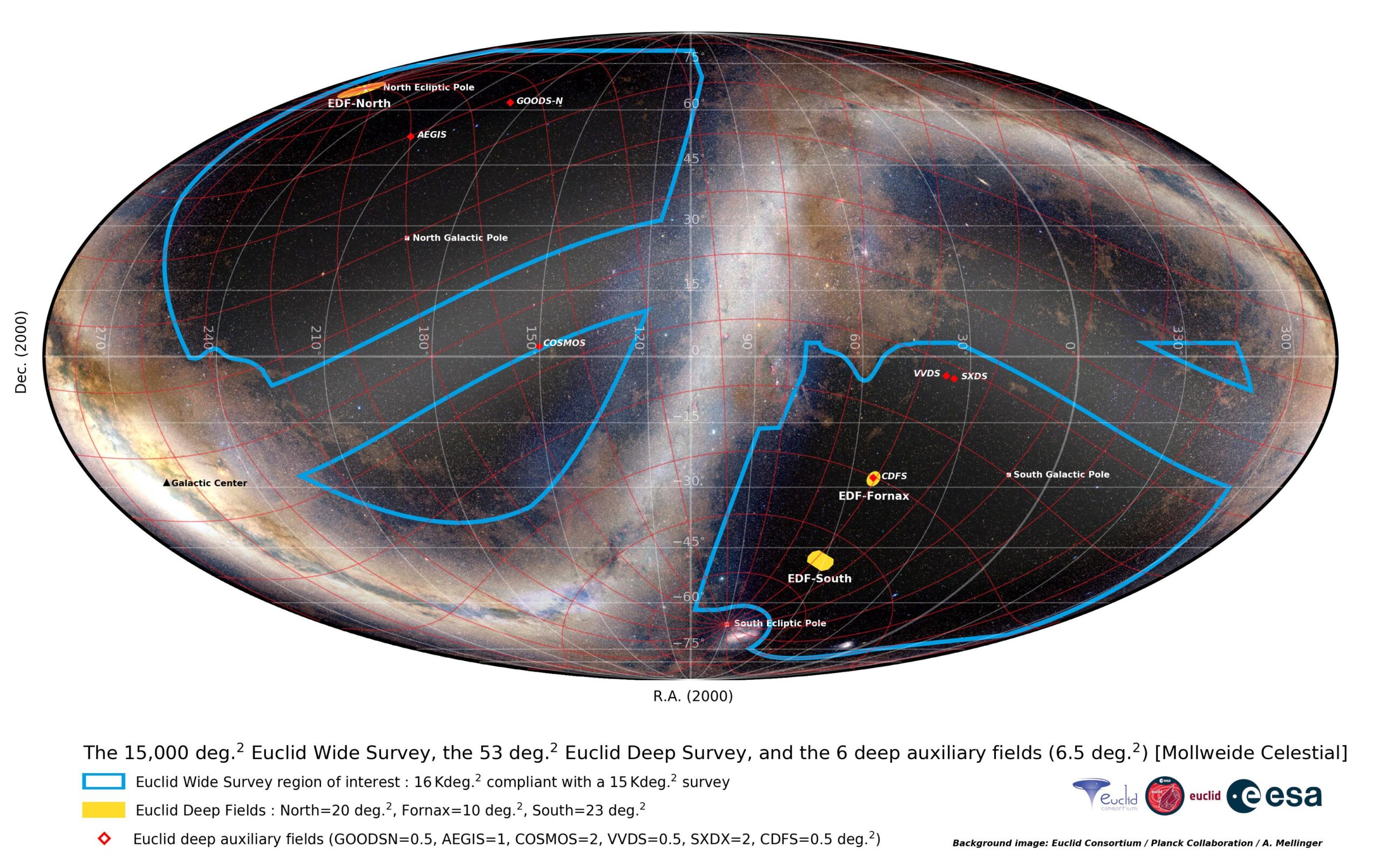}
    \includegraphics[width=18.2truecm]{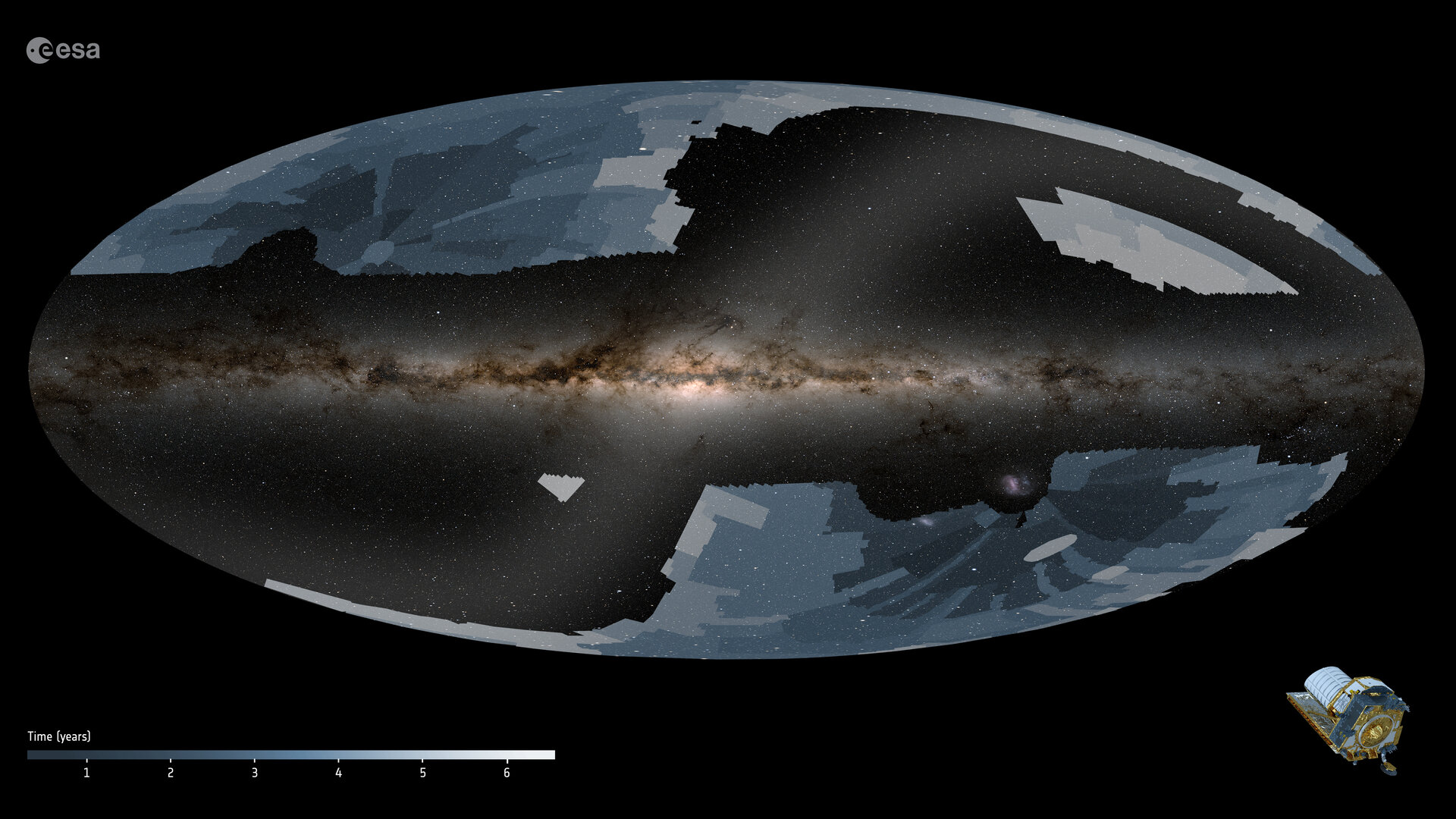}
    \caption{
    \textit{(Top):} Visualization of the \textit{Euclid} survey areas over the sky. 
    \textit{(Bottom):} Survey's planned progression with the operational years. 
    [For panels, credits by: ESA/Euclid\,Consortium/Planck\,Collaboration/A.\,Mellinger.]
    }
    \label{fig:EuclidsSky}
\end{figure*}

On top-panel of Fig.\,\ref{fig:EuclidsSky} is presented the visualization of the \textit{Euclid} survey areas over the sky. 
The bottom-panel instead show the currently planned progression of the survey over the operational years   
[for both panels, credits by: ESA/Euclid\,Consortium/Planck\,Collaboration/A.\,Mellinger.]

\end{appendix}

\end{document}